# Spectral Engineering with CMOS compatible SOI Photonic Molecules

**Luis A. M. Barea,[1,*] Felipe Valini,[1] Guilherme F. M. de Rezende,[1] Newton C. Frateschi[1]**

[1] *Device Research Laboratory, Applied Physics Department, "GlebWataghin" Physics Institute, University of Campinas - UNICAMP, 13083-859 Campinas, SP, Brazil*
[*]*barea@ifi.unicamp.br*

**Abstract:** Photonic systems based on microring resonators have a fundamental constrain given by the strict relationship among free spectral range (FSR), total quality factor ($Q_T$) and resonator size, intrinsically making filter spacing, photonic lifetime and footprint interdependent. Here we break this paradigm employing CMOS compatible Silicon-on-Insulator (SOI) photonic molecules based on coupled multiple ring resonators. The resonance wavelengths and their respective linewidths are controlled by the hybridization of the quasi-orthogonal photonic states. We demonstrate photonic molecules with doublet and triplet resonances with spectral spliting only achievable with single rings orders of magnitude larger in foot print. Besides, these splitting are potentially controllable based on the coupling (bonds) between resonators. Finally, the spatial distribution of the hybrid states allows up to sevenfold $Q_T$ enhancement.

**1. Introduction**

Photonic molecule is the usual name given to two or more electromagnetically coupled optical microcavities designed for spectral engineering, where resonance wavelength and linewidth are tailored independently[1, 2]. The most common types of photonic molecules are based on photonic crystals or ring resonators. While the first approach allows better tunability due to the reduced effective volume, the second allows extra degree of freedom in arranging the multi-coupled-cavity structure and more flexibility in the multiple coupling; footprint reduction and easier scalability/manufacturability[1, 3, 4, 5, 6]. The first applications of photonic molecules were in optical delay line buffers due to their unique dispersion characteristics[4, 7]; optical filters/switches and sensors employing spectral engineering[8, 9, 10, 11, 12, 13] and far-field emission pattern control due to the tuning of the photonic atoms interaction[5, 6, 14]. More recently, Majumdar et. al.[15] proposed to use photonic molecules for quantum simulation because of their improved capacity to generate single photons when one of the cavities contains a strongly coupled quantum dot.

The simplest photonic molecule consists of two coupled ring resonators: analogous to two weakly coupled atoms, it allows a mode splitting and the hybridization of their field distribution at degenerated resonance frequencies. Several new applications have been proposed for this kind of photonic molecule. One is to use coupled cavities (ring resonators or photonic crystals) to increase the four-wave mixing generation efficiency, as did by Azzini et. al.[16]. Another interesting approach is to use their capability of imaginary-frequency resonance splitting to demonstrate a dark state laser, a proposal of Dahlem et.al.[17]. Besides these new applications for microring resonators, such devices are still mainly used as light modulators[18, 19, 20].

In this work, we present planar photonic molecules based on SOI technology and fabricated in a conventional CMOS foundry. The molecule consists of several rings internally coupled to an outer ring coupled to a bus waveguide.  The basic idea of a resonator with internally coupled smaller resonators was proposed by Hong Wei et. al.[21] and Baoqing Su et. al.[22] in a context of Q factor improvement, however never demonstrated experimentally. Here, we demonstrate two building blocks involving one or two ring resonators internally coupled to a larger ring resonator. We show degeneracy splitting and the formation of resonance doublets and triplets as well as a large $Q_T$ enhancement. This enhancement is observed in hybrid eigenmodes that have high photonic density in the small internal rings and low photonic density in the coupling region between the larger outer ring and the bus waveguide. Thus, larger $Q_T$ and larger FSR can be obtained simultaneously. Of particular interest is the interaction between the two inner rings when completely intermediated by the outer ring. This interaction determines the doublet splitting due to the non-orthogonality of the modes, potentially leading to spectral lines separation active control. Therefore, to the best of our knowledge, this is the first demonstration of CMOS compatible planar photonic molecules employing inner coupled ring resonators to obtain degenerate resonance splitting with up to sevenfold $Q_T$ enhancement in a reduced footprint area, that can act as the new building block for many applications.

## 2. Theoretical Analysis

The two SOI photonic molecules with inner coupled rings proposed in this work are shown in Fig. 1. The SOI consists of 220 nm of Si ($n_{Si}$ = 3.45 @ 1550 nm) on 2 µm of $SiO_2$ ($n_{SiO2}$ = 1.45 @ 1550 nm). $SiO_2$ was also used as the top cladding layer covering the devices, with a thickness of 1 µm. To comply with a single-mode condition for transverse electric (TE) polarization at 1.55 µm wavelength, the cross-section dimensions of the waveguides used in this project were 450 nm length and 220 nm height[7, 23]. These dimensions allow low propagation loss and high-index-contrast between the core and cladding layers, giving rise to very strong confinement[24]. This waveguide is coupled to a ring with the same cross-section dimensions and with R = 20 µm. Within this larger ring, a small ring with R = 5 µm is coupled, as shown in Fig. 1(a). Fig. 1(b) shows a similar device but with two coupled cavities within the larger ring.

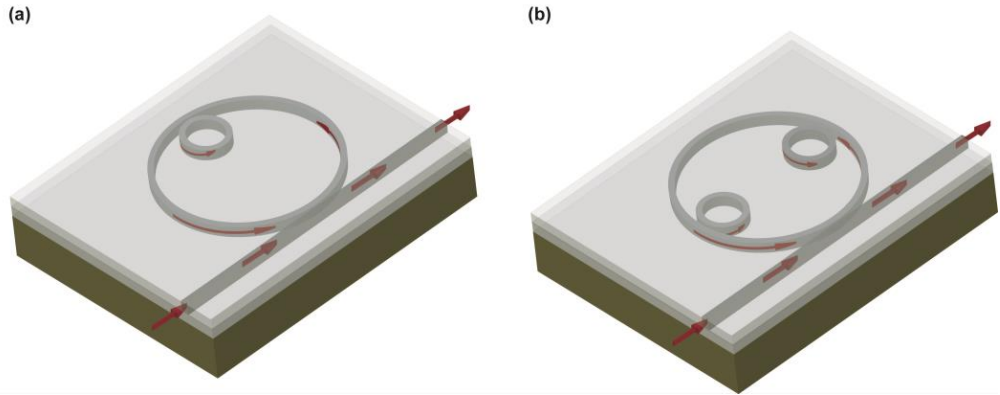

Fig. 1. Schematics of the proposed photonic molecules. (a) One ring coupled to a larger ring and (b) two rings coupled to a larger ring.

The theoretical analysis of these photonic molecules is present in the next two subsections. The first subsection presents simulations using the conventional transfer matrix method (TMM) and Finite-Difference Time-Domain (FDTD). The second subsection shows a brief perturbation theory applied to photonic molecules that allow comprehending the mode splitting / hybridization, giving an insight on the parameters that play an important role on the spectral engineering.

## 2.1 Transmission Spectrum and Modal Analysis

The initial analysis of the transmission spectrum of the photonic molecules was done using the conventional TMM. Eq. (1) is the normalized electric field of output port of the device shown in Fig. 1(a). In this equation, the parameters $\tau_1$ and $\tau_2$ are the transmitting coefficients of the two coupling regions and these parameters are related to the coupling coefficients $k_1$ and $k_2$ by the relation $k_m^2 + \tau_m^2 = 1$, where $m = 1, 2$. The subscript $m = 1$ is associated with the outer ring or waveguide-ring region and $m = 2$ is associated with the inner ring or ring-ring region. From now on, whenever we use the subscript m we are referring to these two situations. $A_m$ is the field attenuation of the outer ring ($m = 1$) and of the inner ring ($m = 2$) for one round trip and they are given by $A_m = e^{-\alpha L_m}$. The parameter α is the loss coefficient of the ring cavity waveguide, ω is the frequency and $L_m$ is the round trip length. $T_m$ is the transit time of one round trip for the rings. These parameters can be written as $T_m = 2\pi N_{eff_m} R_m/c$. $N_{eff}$ is the effective index of the respective microring and this index was connected with the group index ($N_g$) by the expression $N_g = N_{eff} - \lambda(\partial N_{eff}/\partial \lambda)$, where $\lambda$ is the wavelength and $c$ is the light velocity.

$$\frac{E_{out}}{E_{in}} = \frac{\tau_1(1-A_2 e^{iT_2\omega}\tau_2) - A_1 e^{iT_1\omega}(\tau_2 - A_2 e^{iT_2\omega})}{1 - A_2 e^{iT_2\omega}\tau_2 - A_1 e^{iT_1\omega}\tau_1(-A_2 e^{iT_2\omega} + \tau_2)} \tag{1}$$

In our simulations, we consider the especial case of lossless waveguide and the critical coupling condition. To simplify the analysis, we assumed $k_1 = k_2 = 0.1$. The normalized transmission ($|(E_{out}/E_{in})|^2$) is shown in Fig. 2.

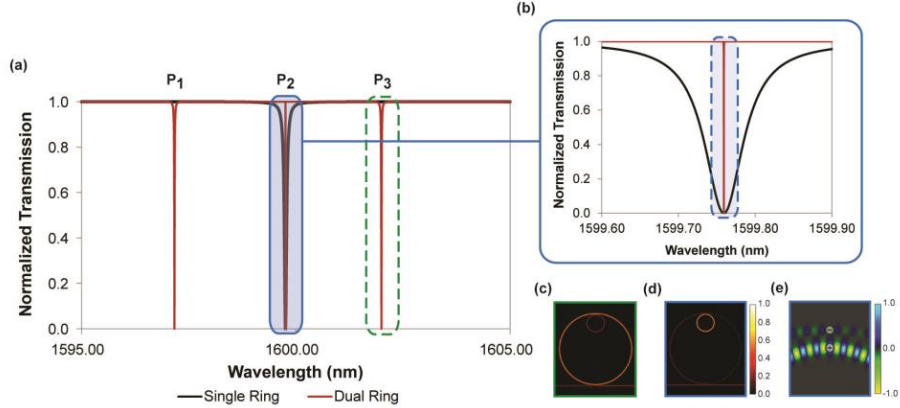

Fig. 2. Simulation results for the photonic molecule with one inner ring: (a) Normalized transmission spectra of output port of dual-ring (red line) and single-ring (black line) resonator. The external ring has R = 20 μm and the internal ring has R = 5 μm. A single ring of R = 5 μm was calculated for reference; (b) Magnified picture of the region where we have the resonance of the internal ring (red line) compared with the resonance of the reference single-ring; (c) and (d) are 2D-FDTD simulations of the power profile at the resonant wavelength of the external ($P_3$) and internal ($P_2$) rings respectively. The resonance $P_2$ is not resonant at the external ring and in (e) we have the electrical field profile at the coupling region between the external ring and the internal ring at this resonance $P_2$. This image shows that the coupling between the rings is anti-symmetric.

The red line in Fig. 2(a) shows the normalized transmission spectrum in a spectral region where an inner ring resonance (Peak $P_2$) lies between two resonances of the larger ring (Peaks $P_1$ and $P_3$). The black line shows the simulation for the case of a single-ring with R = 5 μm coupled to one lossless waveguide. Fig. 2(b) is a zoom in the region near the smaller ring resonance ($P_2$). It is evident that there is a significant narrowing of the linewidth (Δλ) of the smaller ring resonance when it is coupled internally to the larger cavity. The total quality factor ($Q_T = \lambda / \Delta\lambda$) for the single-ring case is about 3 x $10^4$. For the dual-ring situation, $Q_T$ is increased to 2 x $10^7$, i. e., it is improved three orders of magnitude. Fig. 2(c) and 2(d) show 2-D simulations of the power profile at resonances of Fig. 2(a) for the case where the device operates with quasi-transverse electric (quasi-TE) polarization. These simulations were performed by the method of FDTD using the FULLWAVE module of the RSOFT commercial package. The choosing of quasi-TE polarization is because it enables the strongest confinement[24] with minimized loss in bends even for very small radii[25]. As expected, Fig. 2(c) shows that the eigenstate with resonance at 1602.15 nm, a resonance of the larger ring, has essentially all energy stored predominantly in the outer ring. On the other hand, Fig. 2(d) shows that the eigenstate with resonance at 1599.15 nm, a resonance of the smaller ring, is concentrated in the inner cavity. This remote storage of energy in the inner ring allows high $Q_T$ for this cavity. Fig. 2(e) shows 2D-FDTD simulation of the electric field profile in the coupling region between the rings. The symbols plus and minus in this figure represent the maximum and minimum of the electric field, respectively. This result shows that this coupling is anti-symmetric. Moreover, analyzing the peak $P_1$, one can conclude that there is also an increase in $Q_T$ resonance associated with the external cavity. In single-ring case with R = 20 µm, we obtained $Q_T \sim 1$ x $10^5$. The peak $P_1$ of Fig. 2(a) shows a $Q_T \sim 2$ x $10^5$ for the larger ring when the smaller ring is present inside. That is, there are indications that the inner ring resonance filters the outer ring resonance, reducing its $\Delta\lambda$. These results show that for resonances with the field concentrated within the inner ring, $Q_T$ is enhanced with respect to a ring directly connected to the waveguide, because the light cannot populate the larger ring and, therefore, be extracted to the waveguide.

Fig. 1(b) shows a second approach of a photonic molecule with two inner-coupled ring resonators. Eq. (2) is the normalized electric field of the output port for this device.

$$\frac{E_{out}}{E_{in}} = \frac{1}{\tau_1} - \frac{1-\tau_1^2}{\tau_1(1-A_1 e^{iT_1\omega}\tau_1\beta_2\beta_3)} \tag{2}$$

In this Eq. (2), $\beta_2 = \tau_2 - \frac{A_2 e^{iT_2\omega}(1-\tau_2^2)}{1-A_2 e^{iT_2\omega}\tau_2}$ and $\beta_3 = \tau_3 - \frac{A_3 e^{iT_3\omega}(1-\tau_3^2)}{1-A_3 e^{iT_3\omega}\tau_3}$. The subscript 3 refers to the second inner ring coupled to the external ring. Fig. 3 summarizes the simulation results for the degenerate case, where the resonant wavelength of the two small rings and the larger ring is the same. When this condition is satisfied, we expect to see a triplet in the spectrum caused by the hybridization of the three degenerate modes. This condition will be better described in the subsection 2.2, using perturbation theory for photonic molecules. Fig. 3(a) shows the transmission near the degenerate resonance.

A triplet with resonances at $P_{t1}$ = 1540.97 nm, $P_{t2}$ = 1541.24 nm and $P_{t3}$ = 1541.50 nm is observed. Fig. 3(b) shows in detail the triplet present in the spectrum of Fig. 3(a). $Q_T$ for the resonances $P_{t1}$, $P_{t2}$ and $P_{t3}$ are 2 x $10^5$, 2 x $10^7$ and 2 x $10^5$, respectively. These three resonances present in the triplet have $Q_T$ larger than the $Q_T$ for the resonance of a single ring with R = 5 µm ($Q_T$ = 3 x $10^4$) and with similar coupling coefficients. However, the $P_{t2}$ resonance located between $P_{t1}$ and $P_{t3}$, has a $Q_T$ very similar to the case of a single small ring, when the smaller ring resonance lies between the larger ring resonances. Fig 3(c), 3(d) and 3(e) show the results of 2D-FDTD simulation for the power profile in the resonances $P_{t1}$, $P_{t2}$ and $P_{t3}$, respectively, for the case of quasi-TE polarization. The simulation shows that in $P_{t1}$ and $P_{t3}$ resonances, the power is distributed among the three rings. For the case of resonance $P_{t2}$, the result of Figure 3(d) shows that the power is confined in the two inner rings. Therefore, we have again the case where high $Q_T$ is related to a strong confinement within the inner rings and no field in the larger ring. Fig. 4 resumes the electric field profile obtained in the 2D-FDTD simulation for the three cases of the triplet. In this figure, we show the calculated field in the coupling region (rectangles in the schematic drawing) where the symbols $P_{t1}$, $P_{t2}$ and $P_{t3}$ labels the plot for each resonance case. Again, the plus and minus represent the maximum and minimum of the electric field. Analyzing the situation of resonance $P_{t1}$ one can see that the coupling between inner rings and external ring is anti-symmetric, unlike the case of resonance $P_{t3}$, where we have a symmetric coupling. The intensity of the field in the coupling region suggests that for the case of $P_{t1}$ resonance, there is a blue-shift anti-bonding resonance and, for the case of $P_{t3}$ a red-shift bonding resonance. Now, for the case of resonance $P_{t2}$ one can see a case of high hybridization, where we only have high confinement of the field in the inner rings. This result suggests higher energy confinement in the inner rings than the last two cases and this proves the enhancement of the $Q_T$ simulated for this resonance $P_{t2}$.

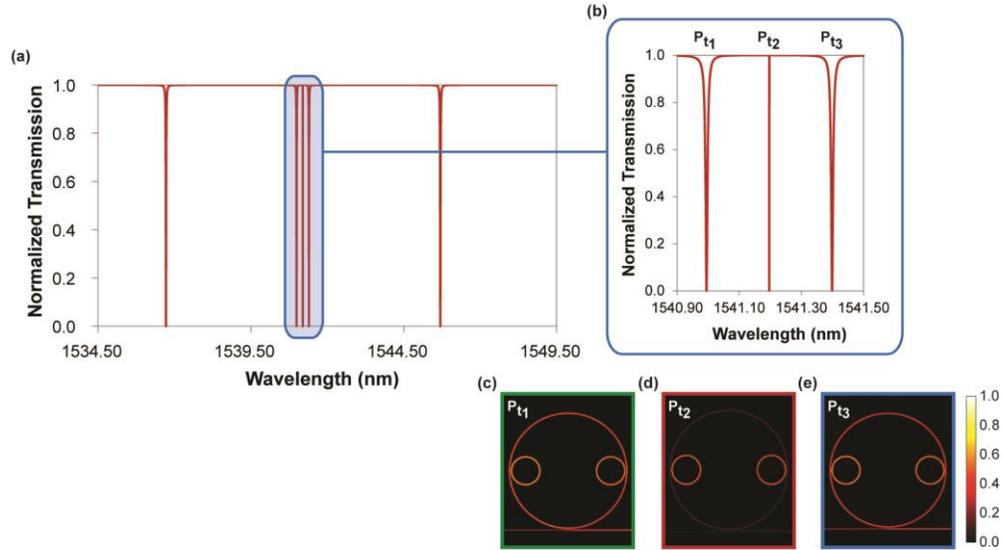

Fig. 3. Simulation results for the photonic molecule with two inner rings (triplet case). (a) Range of the spectrum where it is possible to see a triplet of resonances caused by the hybridization of the three degenerate modes; (b) Magnification of the triplet present in the spectrum of (a); (c), (d) and (e) shows 2D-FDTD simulation for the power profile in the resonances $P_{t1}$, $P_{t2}$ and $P_{t3}$, respectively.

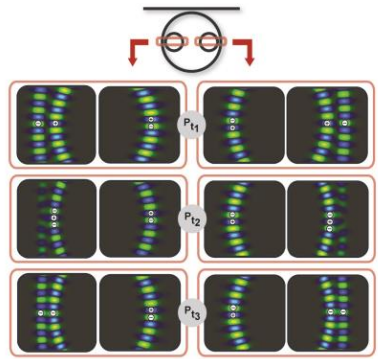

Fig. 4. Electric field profile for the photonic molecule with two inner rings (triplet case). Simulation results obtained in the 2D-FDTD simulation for the three cases of resonances resulted by hybridization of the three degenerated modes. The symbols plus and minus in this figure represent the maximum and minimum of the electric field, respectively.

When these devices operate within a spectral region where the two internal cavities resonances are located between the external cavity resonances, these resonances of the inner cavities suffer a reduction in $\Delta\lambda$ similar to the case shown in Fig. 2(a). However, if the two internal cavities are identical, there is a doublet of resonances, as shown in the results exposed in Fig. 5.

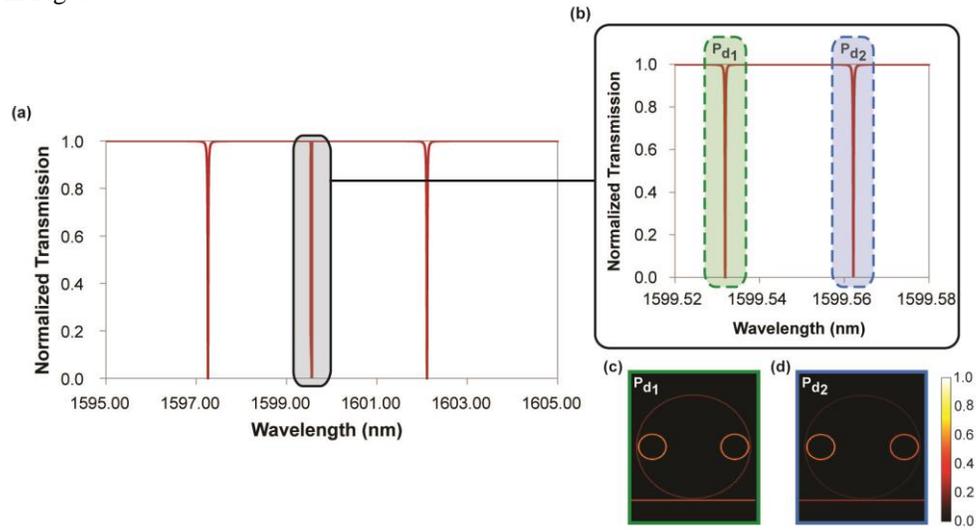

Fig. 5. Simulation results for the photonic molecule with two inner rings (doublet case). (a) Range of the spectrum where it is possible to see a doublet of resonances; (b) Magnification of this doublet present in the spectrum of (a); (c) and (d) show 2D-FDTD simulation for the power profile in the resonances $P_{d1}$ and $P_{d2}$.

This is a typical splitting by anti-crossing of the degenerate states. Fig. 5(a) presents the spectral region where the doublet $P_{d1}$ = 1599.53 nm and $P_{d2}$ = 1599.56 nm are presented. Both resonances are located between the peaks $P_1$ = 1597.26 nm and $P_2$ = 1602.11 nm which are resonant at the external ring. Fig. 5(b) shows in detail this doublet. One can see that the distance between the resonances $P_{d1}$ and $P_{d2}$ is 30 pm and the $Q_T$ of each peak is approximately 1 x $10^7$. Fig. 5(c) and 5(d) are 2D-FDTD simulations of the power profile to $P_{d1}$ and $P_{d2}$ resonances, respectively. One can observe that both cases show that all the power is confined within the two internal cavities. The difference between the two cases can be observed analyzing the phase of the electric field in these two conditions. Fig. 6 shows the electric field profile for the two resonant cases of Fig. 5(b).

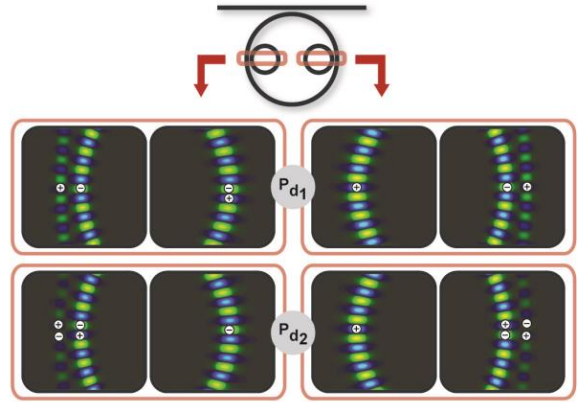

Fig. 6. Electric field profile for the photonic molecule with two inner rings (doublet case): Simulation results obtained in the 2D-FDTD simulation for the case of the doublet of resonances $P_{d1}$ and $P_{d2}$. The symbols plus and minus in this figure represent the maximum and minimum of the electric field, respectively.

Analyzing Fig. 6, one can see that for both situations the inner rings are anti-symmetrically coupled to the external ring, but there is a phase-shift between these two cases. Furthermore, in both cases the intensity of the field is much higher in the inner rings that in the external ring. Again, this suggests that we have energy stored inside both inner rings, increasing the $Q_T$ of these resonances. However, these simulations does not explain how is possible to have a doublet since the two inner resonators are not coupled to each other. So, we developed a perturbation theory applied to these photonic molecules to explain how is the coupling between these two inner rings, which in the end is somehow mediated by the outer ring.

*2.2 Perturbation Theory Applied to Photonic Molecules*

The essential features described by the TMM or FDTD approaches can also be obtained using the following simple perturbation theory, much similar to molecules based on weakly bonded atoms. In this approach, the transversal electromagnetic field component confined in a microring resonator can be described by the state $|m, l\rangle_0$, where m is the mth resonator of the coupled system and $l$ represents the light azimuthally propagation direction related to the photon angular momentum ($l = +$ stands for clockwise and $l = -$ for counterclockwise). Assuming $|m, l\rangle_0$ stands for an unperturbed electric field component of the mth microring and that the m microrings of the system have a degenerated eigenfrequency $\omega_0$, the solution of the Helmholtz equation follow as:

$$\left[\nabla_0^2 + \left(\frac{n\omega_0}{c}\right)^2\right]|m, l\rangle_0 = 0 \Rightarrow \nabla_0^2 |m, l\rangle_0 = -\left(\frac{n\omega_0}{c}\right)^2 |m, l\rangle_0 \qquad (3)$$

where n is the refractive index and c the speed of light.

The proximity of one resonator to another inserted into the system causes a perturbation of the refractive index in the form $n + \Delta n$, where $\Delta n \ll n$ so that $(n + \Delta n)^2 \approx n^2 + 2\,n\,\Delta n$ (neglecting higher orders terms). The perturbed field $|m, l\rangle$ satisfies:

$$\left[\nabla^2 + \frac{2\,n\,\Delta n\,\omega^2}{c^2}\right]|m, l\rangle = -\left(\frac{n\omega}{c}\right)^2 |m, l\rangle \qquad (4)$$

Although the unperturbed confined modes do not form a complete set of orthogonal functions, we propose that approximately one can express the perturbed field $|m, l\rangle$ as a superposition of the unperturbed modal fields:

$$|m, l\rangle = \sum_{m,l} a_m |m, l\rangle_0 \qquad (5)$$

Substituting the proposed solution Eq. (5) into the wave equation Eq. (4), multiplying by $_0\langle i, z|$ and integrating over all space one can obtain:

$$a_i(\omega - \omega_0) + \sum_{j,k} a_{i,l} \,_0\langle i, k| \frac{\Delta n\,\omega_0}{n}|m, l\rangle_0 \approx 0 \qquad (6)$$

To obtain Eq. (6) we assumed that $_0\langle i, k|m, l\rangle_0 = \delta_{i,m}\,\delta_{k,l}$, which in principle is not true because microrings are always dissipative systems with complex eigenfrequencies $\omega$. Nevertheless, $\Delta n$ is very small and it is presented in such a small region of the structure (coupling region) that makes our assumption valid. We then rewrite the linearized Eq. (6) as:

$$a_i(\omega - \omega_0)\frac{1}{\omega_0} + \sum_{m,l} a_{m,l} W_{il,mk} = 0 \qquad (7)$$

where $W_{il,mk}$ is the perturbation matrix:

$$W_{il,mk} = \,_0\langle i, l|\frac{\Delta n}{n}|m, k\rangle_0 \qquad (8)$$

To understand each element of $W$, it is necessary to understand how the ith microring interacts with the jth microring. Defining $\phi$ as the angular coordinate of the ith resonator, whose origin stands at the closest distance between the microrings, we can illustrate the refractive index perturbation dependence with $\phi$ in Fig. 7 (a). The schematic, that indicates the index of each microrings used in the perturbation theory for photonic molecules, is present in Fig. 7(b).

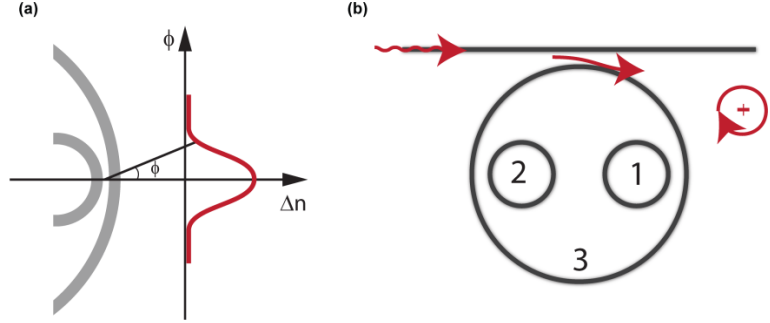

Fig. 7. Refraction index perturbation profile and schematic of device with three rings coupled. (a) Azimuthal dependence ($\phi$) of the refraction index perturbation near the coupling region between the microrings; (b) The schematics indicates the index of the microrings used in the perturbation theory for photonic molecules.

Now, the general matrix element can be written as:

$$W_{ml,m'k} = \frac{1}{2\pi R n} \int_0^{2\pi} e^{ilm\phi} e^{-ikm'\phi} R\, \Delta n(\phi) d\phi \qquad (9)$$

For the $W_{ml,ml}$ elements, considering that $\Delta n(\phi)$ is randomly distributed over each ring one can see that:

$$W_{ml,ml} = \frac{1}{2\pi n} \int_0^{2\pi} \Delta n(\phi) d\phi \approx 0 \qquad (10)$$

For now on, we use the schematics of Fig. 7(b), labeling the microrings as 1, 2 and 3. A reasonable approach to evaluate the other matrix elements (assuming the inner rings have equal dimensions) is to consider that the coupling between the outer ring with each one of the inner rings is possible only if the light is propagating in the same direction:

$$W_{3l,1k} = W_{3l,2k} \propto \delta_{l,k} \Rightarrow W_{3l,1k} = W_{3l,2k} \equiv b\, \delta_{l,k} \qquad (11)$$

where $b$ is the coupling factor for this case. If we consider that roughness can couple counter propagating modes (with opposite $l$), there will be a coupling element defined by r. If the dimensions of the inner rings are increased, allowing the evanescent field of the inner rings to overlap, arises a new coupling term c between opposite $l$'s for the two inner rings:

$$r \equiv W_{i+,i-} \qquad (12\text{-}1)$$
$$c \equiv W_{1+,2-} \qquad (12\text{-}2)$$

Moreover, we find necessary to introduce a coupling coefficient to explain the doublet formation evidenced in Fig. 5:

$$\bar{b} \equiv W_{1l,2l} \qquad (12\text{-}3)$$

This coefficient couples the inner rings if they have the same $l$, even if the evanescent modes of them do not overlap ($c = 0$), which means this coupling is intermediated by the outer ring. If one appeal to the standard coupling mode theory, such assumption is easily explained considering an oscillator that intermediates a coupling between two independent oscillators.

Finally, since our system is symmetric and assuming that all the elements are real, we can define the other terms in the same way and construct the entire 6 x 6 eigenvalue matrix:

$$
\begin{array}{c|cccccc}
 & |1+\rangle & |2+\rangle & |3+\rangle & |1-\rangle & |2-\rangle & |3-\rangle \\
\langle 1+| & \frac{(\omega-\omega_0)}{\omega_0} & \bar{b} & b & r & c & 0 \\
\langle 2+| & \bar{b} & \frac{(\omega-\omega_0)}{\omega_0} & b & c & r & 0 \\
\langle 3+| & b & b & \frac{(\omega-\omega_0)}{\omega_0} & 0 & 0 & r \\
\langle 1-| & r & c & 0 & \frac{(\omega-\omega_0)}{\omega_0} & \bar{b} & b \\
\langle 2-| & c & r & 0 & \bar{b} & \frac{(\omega-\omega_0)}{\omega_0} & b \\
\langle 3-| & 0 & 0 & r & b & b & \frac{(\omega-\omega_0)}{\omega_0}
\end{array}
\qquad (13)
$$

where $\frac{(\omega-\omega_0)}{\omega_0}$ are the eigenvalues we are interested to obtain. Since the diagonal elements of the perturbation $W$ are nulls it is immediate to note that $\text{Tr}(W) = 0$. A corollary is that the sum of the eigenvalues of the system is zero. Now the system can be applied to solve each one of our photonic molecules.

### 2.2.1  One inner ring coupled to an outer ring

In the case of Fig. 2(a), we are interested in the resonances of the inner ring that are not resonant with the outer ring. Thus, eliminating the outer ring, the matrix of this system is the Eq. (13) reduced to:

$$
\begin{array}{c|cc}
 & |1+\rangle & |1-\rangle \\
\langle 1+| & \frac{(\omega-\omega_0)}{\omega_0} & r \\
\langle 1-| & r & \frac{(\omega-\omega_0)}{\omega_0}
\end{array}
\qquad (14)
$$

Since no splitting is observed in the resonance transmission spectrum of Fig. 2(a) one we can conclude that $r = 0$ in the reduced matrix of Eq. (14). So from now on we will not include the effects of roughness anymore.

### 2.2.2  Two inner rings coupled to an outer ring

Now, consider the situation of two inner rings coupled to an outer ring as shown in Fig. 1(b). For this specific photonic molecule, we have two cases to analyze with our perturbation method: the emergence of doublets and triplets. The only difference about this photonic molecule with respect to the previous one discussed before is an extra inner ring, so it is not expected any modification of the coefficient r which was proved to be null. Since the two inner rings are not directly coupled, the coefficient c is also null. The matrix Eq. (13) is now written as:

|           | $|1+\rangle$                        | $|2+\rangle$                        | $|3+\rangle$                        | $|1-\rangle$                        | $|2-\rangle$                        | $|3-\rangle$                        |
|-----------|-------------------------------------|-------------------------------------|-------------------------------------|-------------------------------------|-------------------------------------|-------------------------------------|
| $\langle 1+|$ | $\frac{(\omega-\omega_0)}{\omega_0}$ | $\bar{b}$                           | $b$                                 | 0                                   | 0                                   | 0                                   |
| $\langle 2+|$ | $\bar{b}$                           | $\frac{(\omega-\omega_0)}{\omega_0}$ | $b$                                 | 0                                   | 0                                   | 0                                   |
| $\langle 3+|$ | $b$                                 | $b$                                 | $\frac{(\omega-\omega_0)}{\omega_0}$ | 0                                   | 0                                   | 0                                   |
| $\langle 1-|$ | 0                                   | 0                                   | 0                                   | $\frac{(\omega-\omega_0)}{\omega_0}$ | $\bar{b}$                           | $b$                                 |
| $\langle 2-|$ | 0                                   | 0                                   | 0                                   | $\bar{b}$                           | $\frac{(\omega-\omega_0)}{\omega_0}$ | $b$                                 |
| $\langle 3-|$ | 0                                   | 0                                   | 0                                   | $b$                                 | $b$                                 | $\frac{(\omega-\omega_0)}{\omega_0}$ |

(15)

The system can be decoupled into two subsystems for the light propagating independently in each one of the two possible directions, clockwise and counter clockwise respectively (+ or -). So, after reducing the matrix, three eigenvalues can be obtained:

$$\frac{(\omega-\omega_0)}{\omega_0} = -\bar{b} \tag{16-1}$$

$$\frac{(\omega-\omega_0)}{\omega_0} = \frac{1}{2}\left(\bar{b} - \sqrt{8b^2 + \bar{b}^2}\right) \tag{16-2}$$

$$\frac{(\omega-\omega_0)}{\omega_0} = \frac{1}{2}\left(\bar{b} + \sqrt{8b^2 + \bar{b}^2}\right) \tag{16-3}$$

These three eigenvalues explain the triplets along the transmission spectrum of Fig. 3. The first frequency of Eq. (16) is for the case where the inner rings are coupled to the outer ring and the second and third frequencies are for the case where the inner rings have a degenerated eigenstate uncoupled of the outer ring.

Considering now the last case, where the inner rings resonances are not resonant with the outer ring, as shown on Fig. 5, we can reduce the matrix of Eq. (15) to:

|           | $|1+\rangle$                        | $|2+\rangle$                        | $|1-\rangle$                        | $|2-\rangle$                        |
|-----------|-------------------------------------|-------------------------------------|-------------------------------------|-------------------------------------|
| $\langle 1+|$ | $\frac{(\omega-\omega_0)}{\omega_0}$ | $\bar{b}$                           | 0                                   | 0                                   |
| $\langle 2+|$ | $\bar{b}$                           | $\frac{(\omega-\omega_0)}{\omega_0}$ | 0                                   | 0                                   |
| $\langle 1-|$ | 0                                   | 0                                   | $\frac{(\omega-\omega_0)}{\omega_0}$ | $\bar{b}$                           |
| $\langle 2-|$ | 0                                   | 0                                   | $\bar{b}$                           | $\frac{(\omega-\omega_0)}{\omega_0}$ |

(S17)

Again, we can decouple the clockwise and counterclockwise light way propagation to obtain the two degenerated eigenfrequencies that explain the presence of the doublet:

$$\frac{(\omega-\omega_0)}{\omega_0} = +\bar{b} \tag{18-1}$$

$$\frac{(\omega-\omega_0)}{\omega_0} = -\bar{b} \tag{18-2}$$

Here we have a typical result of anti-bonding and bonding that result in a splitting analogous to electronic states coupling, which gives origin to the term photonic molecules.

One should note that we have a splitting only if $\bar{b}$ is considered, confirming that somehow the outer ring must intermediate the coupling between the two inner rings. Further in this work we will demonstrate experimentally that this indeed occurs. This effect may be better investigated using an approach involving a multiple coupled oscillators that is a matter of future publication. This opens a possibility of splitting active control.

## 3. Device design and fabrication

The devices were fabricated using an SOI platform at IMEC-EUROPRACTICE, using design rules fully compatible with CMOS process. The final devices are illustrated inset in Fig. 8(a) and 9(a). These images show an optical microscope micrograph of two devices used as proof of concept of the inner ring resonators coupling approach. These devices consist of typical SOI all-pass microring resonators. In both images the larger microring resonator has R = 20 µm and is coupled to a waveguide by a 200 nm gap. These waveguides have a core cross-section of 450 nm x 220 nm. The device of inset in Fig. 8(a) shows a 5 µm radius ring internally coupled to the larger microring and the figure inset in Fig. 9(a) has two 5 µm radius microring resonators internally coupled to the larger ring. The internal coupling also employs a 200 nm gap. The chips containing these devices have been spliced in areas where there are inverse nanotapers for efficient coupling to optical fibers[26]. In order to obtain high-quality mirrors with excellent coupling, we used a polishing technique with focused ion beam (FIB) milling[27]. The polishing technique consists of milling the input and output ports where the nanotapers are located. $Ga^+$ ion beam at 30 kV and an emission current of 0.1 nA were used. Single rings with R = 20 µm and R = 5 µm coupled only to a bus waveguide with 200 nm gap were also fabricated for reference.

## 4. Experimental Results

The optical measurements of the transmission spectra were performed with nano-positioners to align polarization controlled grin rod lensed optical fibers to the bus waveguide at the input and output ports. A tunable laser source with wavelength ranging from 1465 nm to 1630 nm was used as light source and a fiber-coupled power meter was used to measure transmitted signals. A Peltier cooler set at 20°C was employed for all measurements.

    The transmission spectrum of the fabricated device with one inner ring is shown in Fig. 8(a). One resonance of the inner ring at 1599.16 nm is observed and this resonance is located between two resonances of the larger ring, 1597.16 nm and 1602.11 nm. In Fig. 8(b) this inner ring resonance is shown in detail (black dots). The blue dots are the transmission spectrum of a typical 5 µm radius single ring directly coupled to a waveguide with a gap of 200 nm. This single-ring device was fabricated in the same chip for reference. The linewidth ($\Delta\lambda$) of the resonance of the inner 5 µm radius ring (Lorentzian fit - red line) is 40 pm. This $\Delta\lambda$ corresponds to a $Q_T$ of about 40,000. In comparison, the reference single ring (R = 5 µm) has a $Q_T$ of 8,000, i. e., this dual-ring device really enables the increase of five times in $Q_T$ with reduced footprint area (40 µm x 40 µm). One may argue this comparison is inexact, however, it should be noticed that this photonic molecule also has a sevenfold enhancement with respect to a single ring of 20 µm radius, and featuring an extra resonance. Fig. 8(c) is a false colored infrared image when the photonic molecule is pumped at the resonant wavelength of the inner ring ($\lambda$ = 1599.16 nm). One can observe that light is concentrated just in the inner ring since it is off-resonance in the larger ring. This result is similar to the power profile presented in the simulation result of Fig. 2(d). The increase in $Q_T$ can be explained by the relative decoupling between the smaller ring and the bus waveguide caused by the off-resonance of the larger ring. Previous simulation results showed that the coupling between the two cavities in this case is anti-symmetric and the electric field in the external ring is very low compared with the inner ring. Therefore, this condition shows that using a photonic molecule with two coupled rings, it is possible to store great part of the energy inside the inner ring, causing a fivefold enhancement in the $Q_T$, with reduced FSR compared to the single ring case and with small footprint area. Also, one must observe that the degeneracy between clockwise and counterclockwise modes is not broken. In the subsection 2.2.1, we neglect the coupling term

between counter-propagating modes in the same ring. The red line showed in Fig. 8(a) is the simulated curve for the dual-ring device using Eq. (1). With this fitting, we can obtain that the larger ring is coupled to the bus-waveguide with a coupling factor $k_1 \sim 0.4$ and the inner ring is coupled to the external ring with a coupling factor $k_2 \sim 0.4$. Therefore, if one performs the same simulation fitting with the same experimental parameters but assuming $k_1 = k_2 = 0.1$, an experimental $Q_T = 380,000$ for the inner ring resonance is obtained. This shows that the transmission spectrum and $Q_T$ of the inner ring is very sensitive to the coupling factor and can reach a high $Q_T$ if the critical coupling can be tuned with more accuracy.

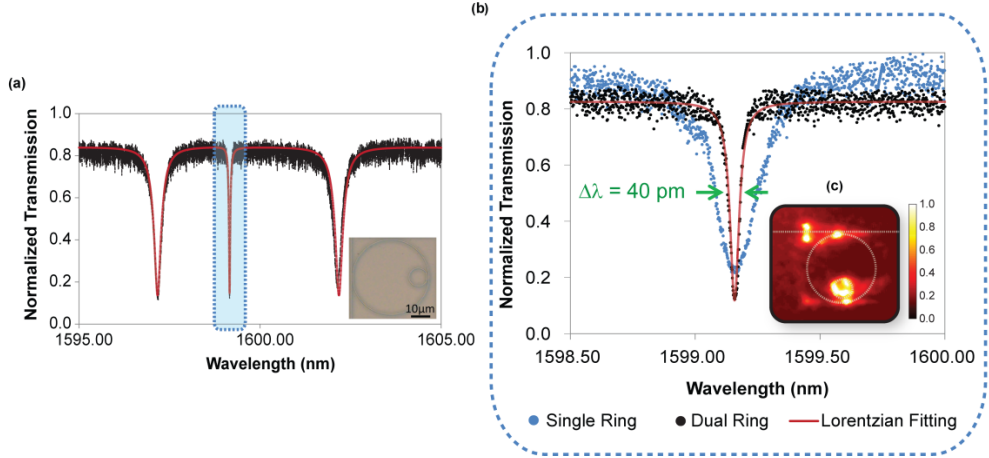

Fig. 8. Photonic molecule with one inner ring. (a) Transmission spectrum of the device with one inner microring. The solid red line is a simulated fitting using Eq. (1). The image inset shows the device fabricated with one inner ring coupled to the external ring; (b) Comparison between the resonance of inner microring (black dots) and one single microring resonator (R = 5 μm - blue dots) directly coupled to a waveguide. The solid red line in this figure is a Lorentzian fitting of the inner-ring resonance; (c) Infrared image (false colored) taken at the resonance of the inner cavity.

Similar to atomic coupling, if one can devise a combination of ring resonators with degenerate modes, one would expect, upon hybridization, frequency splitting. This splitting may favor a particular energy spatial distribution that may lead to high $Q_T$, if the mode hybridization results in low field intensity near the coupling regions to the bus waveguide. This is the case of two inner rings coupled to an outer ring. Fig. 9(a) shows the transmission spectra of the fabricated photonic molecule with two inner microring resonators, as shown inset in Fig. 9(a). The insertion of a second inner ring results in threefold degeneracy between the large ring and the two small rings and this mode hybridization should result in a triplet splitting. Such coupling condition is showed in Fig. 9(a), where we have the transmission spectrum with a triplet of resonances at $P_{t1} = 1540.80$ nm, $P_{t2} = 1541.25$ nm and $P_{t3} = 1541.65$ nm. The red line in this figure shows the simulation fitting using Eq. (2). The fitting suggests that the coupling factors are approximately $k_2 = k_3 = 0.2$, near $\lambda = 1541.20$ nm. Fig. 9(b) shows in detail the triplet presents in the spectrum of Fig. 9(a) and one can see that the $Q_T$ for the resonance $P_{t1}$, $P_{t2}$ and $P_{t3}$ are 16,000, 43,000 and 19,000, respectively. This $Q_T$ was calculated using a Lorentzian fitting represented by the blue, green and yellow curves, respectively. As suggested by the simulations, the triplet resonances have larger $Q_T$ compared to the resonance of only a single ring with R = 5 μm. This single ring presented $Q_T \sim 6,000$ at $\lambda = 1541.88$ nm. For the single ring case with R = 20 μm also fabricated in the same chip, we obtained a $Q_T \sim 10,000$ at $\lambda = 1541.25$ nm. If we also compare this $Q_T$ with each value for the triplet case, one can conclude that the $Q_T$'s of the triplet resonances are larger than the case where we have a single-ring with R = 20 μm. Compared to the single-ring (R = 5 μm), one can show that $Q_T$ is increased almost three times for the resonances $P_{t1}$ and $P_{t3}$. For the $P_{t2}$ resonance, the $Q_T$ is increased almost 7 times with respect to the single ring and this $Q_T$ is higher than the narrow resonance measured in Fig. 8(b) for the dual-ring system. Fig 9(c), 9(d) and 9(e) show infrared images (false colored) of the device pumped at the resonances $P_{t1}$, $P_{t2}$ and $P_{t3}$, respectively. These images can be compared with the FDTD simulations of Fig. 3(c), 3(d) and 3(e) in section 2.1. Although the image is not as clear as the simulation, one can observe that the infrared images shown in Fig. 9(c) and 9(e) show the power distributed among the three resonators for $P_{t1}$ and $P_{t3}$ resonances. Note that only in these two images one can see the external ring with some light. Fig. 4 in the simulation results (section 2.1) suggested that for the case of $P_{t1}$ resonance there is a blue-shift anti-bonding resonance and for the case of $P_{t3}$ a red-shift bonding resonance. For the resonance $P_{t2}$, the infrared image shown in Figure 9(d) exhibits the stored energy essentially distributed within the two inner rings and the simulation of Fig. 3(d) and Fig. 4 show that this resonance $P_{t2}$ is a case of hybridization, where light is confined mainly in the inner rings. Therefore, similar to the off resonant case that we described for the single inner ring, the hybridization reduces the field intensity in the coupling region to the bus waveguide increasing the $Q_T$ in a similar manner as dark states[28, 29].

Once more, after an analogy to an atomic system, the number of resonance splitting that one can obtain is equal to the degeneracy of the system, as shown in section 2.2. Increasing even more the number of inner ring resonators is possible to create a photonic molecule with tailored spectra in a very small footprint area.

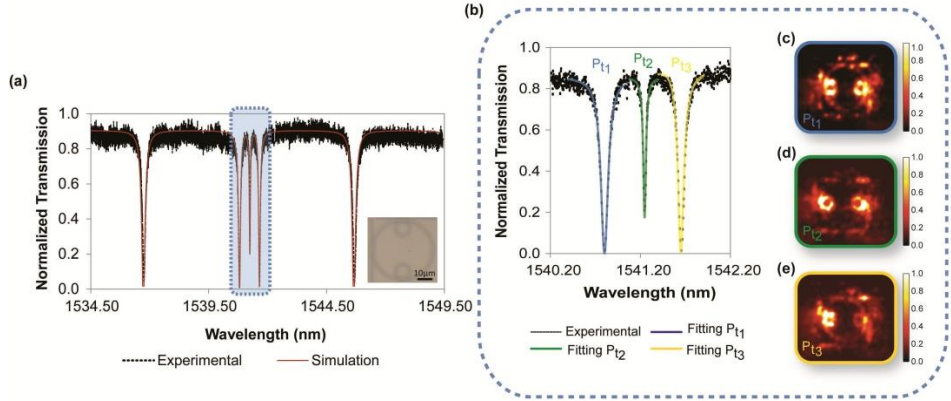

Fig. 9. Photonic molecule with two inner rings (triplet case). (a) Transmission spectrum of the photonic molecule with two inner rings within a spectrum region where it is possible to see a triplet of resonances. Figure inset show the photonic molecule fabricated with two inner rings coupled to an external ring; (b) Details of the triplet caused by the hybridization of the three degenerated modes of the inner and external rings; (c), (d) and (e) show infrared images (false colored) of the device tunable in the three resonances of the triplet.

Analyzing a different region of the same spectrum of this photonic molecule, shown in Fig. 10(a), one sees that the transmission presents two narrow resonances at $\lambda_{Pd1}$ = 1598.57 nm and $\lambda_{Pd2}$ = 1598.91 nm which are not resonant in the larger ring. The presence of these two resonances represents a splitting of the two degenerate coupled resonators, exactly similar to the case shown in the simulation results in Fig. 5. It is very important to note that this splitting is not caused by a fabrication error that leads to two slightly different rings. If this was the case, a 4 fold splitting and not 3 fold splitting would be observed in the above case where the outer and the two inner rings are degenerate. Therefore, the doublet is indeed occurring due to an interaction between the two inner disks even though they do not interact directly. The perturbation theory applied to this photonic molecule confirms the existence of triplets and doublets. More than that, it shows that the doublet can only exist if the inner ring couplings are off-resonance with the outer ring but somehow has the coupling between them intermediated by the outer ring. This is indeed possible because the photonic states are not orthogonal and there is always coupling between the modes. The wavelength splitting between these two resonances is 340 pm and this value is related to the coupling factor between these two cavities. The red line is the fitting using Eq. (2) and this result suggests that the inner rings were coupled to the external ring with coupling factor of approximately $k_2 = k_3 = 0.3$. Fig. 10(b) is the magnification of the resonances related to the inner rings. This figure shows that we have higher $Q_T$'s for both resonances, $Q_{Pd1} \sim 23,000$ and $Q_{Pd2} \sim 37,000$. Once more, the $Q_T$ of all inner ring resonances increased when off resonance in the outer ring. Compared with the case where we have single ring resonator (R = 5 μm) with $Q_T \sim 8,000$, these measured values represent an increase of almost 3 times for the $P_{d1}$ resonance and 5 times for the $P_{d2}$ resonance. The reduction in $Q_{Pd1}$ compared with the $Q_{Pd2}$ value occurs because the resonance $P_{d1}$ is affect by roughness. This is confirmed by the fact that a small splitting is observed in this mode, as shown with a zoom in this peak. In this case, the counter propagating waves coupling factor $r$ described in the section 2.2 can no longer be neglected. It is interesting to note that this photonic molecule in this condition is analogous to the case where an Electromagnetic Induced Transparency (EIT) phenomenon occurs[30]. In our case, the two inner rings represent the dressed levels of a split atomic level while the outer ring represents the pump from an external level. In EIT, the absorption is drastically reduced due to a coherent optical nonlinearity arising from the strong coupling of photons with matter. In our case, we simply increase the photon lifetime inside the two inner rings allowing a very narrow window of transparency.

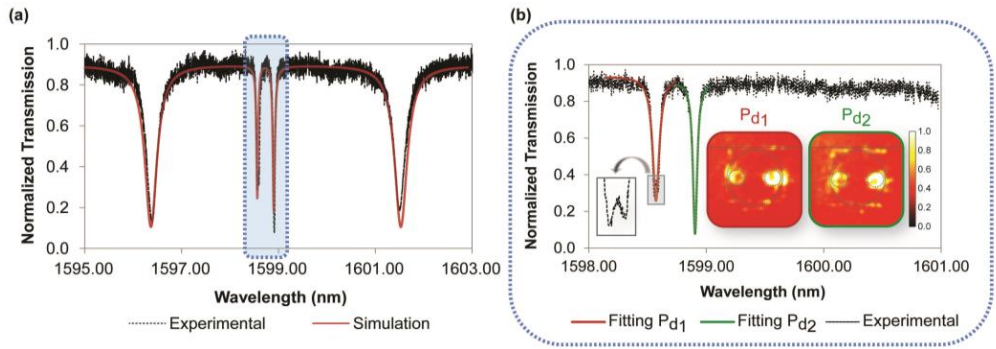

Fig. 10. Photonic molecule with two inner rings (doublet case). (a) Transmission spectrum of the photonic molecule with two inner rings; (b) Detail of the doublet caused by the two degenerated resonances associated with the two inner rings. The red and green solid lines are Lorentzian fittings of the two resonances and a zoom of the splitting in $P_{d1}$ is shown in detail. The inset figures show infrared images (false colored) of the two resonances of this doublet.

The figures in the inset at Fig. 10(b) show infrared images of the cavity pumped exactly at each resonant wavelength ($P_{d1}$ and $P_{d2}$). One can compare this image with the FDTD simulation shown in Fig. 5(c) and 5(d). This comparison suggests that we really have the energy stored inside both inner cavities in this condition of a doublet splitting. Moreover, one can see in the simulation of Fig. 6 that the coupling between the inner rings and external ring is anti-symmetric in both resonances, differing only by a shift in phase. Furthermore, this coupling between the inner rings is mediated by a lossy medium that is the external ring resonator and this coupling can be tuned through the external ring. For example, we may possibly change the coupling of the internal cavities by changing the temperature of the external ring[31]. Thus, one can control the splitting without directly acting on the internal resonators.

## 5. Conclusion

In summary, we demonstrated two CMOS compatible photonic molecules based on internally coupled microring resonators that break the paradigm of the interdependence between photonic lifetime, spectral spacing and footprint. The first photonic molecule, with only one inner microring, has a fivefold $Q_T$ enhancement and reduced footprint when the resonance of the inner ring is between two resonances of the outer ring. This device can be very useful as a sensor since high quality factor can be obtained for very small rings.

The second photonic molecule, with two inner microrings, also shows sevenfold $Q_T$ enhancement, and present resonance splitting, analogous to a weakly coupled atomic molecule. One of the splitting is a doublet with a sevenfold $Q_T$ enhancement with approximately 40 GHz of separation. Also, in this same photonic molecule, we presented a triplet splitting, with two threefold and one sevenfold $Q_T$ enhancement and separation of order of 50 GHz.

Very compact photonic molecules with the potential for tuning the coupling factors and spectral splitting via temperature and / or carrier injection can be very useful for integrated optical processing. For instance, one can use them to filter sidebands of modulated optical signals and generated RF signals. Also it is possible to store the carrier and the sidebands of a modulated optical signal together, if one use the triplet splitting. With the high power confined into the inner microrings one can obtain enhanced nonlinear effects in a reduced footprint. A last approach is to insert a gain media into the inner microring, allowing the integration of a coherent light source with Silicon photonics. Therefore, this class of photonic molecules may become the new building blocks for CMOS compatible and foundry-manufacturable Silicon photonics.

**Acknowledgment**

The authors would like to thanks Thiago Alegre and Gustavo Wiederhecker for helpful discussions. This work was supported by the Brazilian financial agencies: CNPq, CAPES, Center for Optics and Photonics (CePOF) under grant # 05/51689-2 and National Institute for Science and Technology (FOTONICOM) under grant #08 /57857-2, São Paulo Research Foundation (FAPESP).